          [1999/12/02 1.01 (PWD)]
\documentclass[a4paper,twocolumn]{esapub} % European paper
\usepackage{natbib}
\usepackage{graphicx}

\title{X-ray transients in the galactic center region}
\author{Jean in 't Zand}
\affil[]{Space Research Organization Netherlands, Utrecht}

\def\lum{erg~s$^{-1}$}
\begin{document}

\keywords{X-ray binaries, X-ray bursts, neutron stars}

\raggedbottom
\maketitle

\begin{abstract}
In the last 15 years, 6 dedicated observation programs were carried out
to monitor $\sim20^{\rm o}$ around the galactic center for
transient activity above a few keV. Transient activity from low-mass
X-ray binaries shows a strong preference for this field. Two programs are
currently active, with the Wide Field Cameras on BeppoSAX and the
Proportional Counter Array on RXTE. The coverage of these programs is fairly
extensive, with typical sensitivities of a few mCrab and an angular
resolution of a few arcminutes which is better than that of typical all-sky
monitor devices. Fifteen transients with peak fluxes above 10 mCrab were
discovered with these instruments so far (equivalent to 12$\pm4$ such
transients per year over the whole sky), and on top of that BeppoSAX-WFC
detected about 2000 X-ray bursts. We summarize some recent results.
\end{abstract}

\section{Introduction}

The INTEGRAL mission will dedicate half of its Core Program to 
semi-annual deep exposures of the central radian the Galaxy,
and a quarter to weekly scans of the galactic plane (Winkler 
2000). Together with incidental observations from the rest of
the Core Program and the General Program, the coverage of the
galactic center region will be large. In this respect it is
interesting to assess what prospects lie ahead for INTEGRAL in
the area of detecting X-ray transients. Thanks to dedicated and
sensitive observations with BeppoSAX and RXTE in recent years
the observational database about these transients is growing.
We summarize recent developments.

\section{Nature of transients}

We define a transient as a source that in the medium X-ray energy
range (a few keV to a few tens of keV)
\begin{itemize}
\itemsep=0cm
\item has a quiescent level below 1 mCrab (equivalent to $\sim10^{35}$~\lum\
      if at a distance equal to that of the galactic center);
\item if it peaks above 10 mCrab;
\item if it shows activity for at least 10~sec;
\item if it is mostly in quiescence.
\end{itemize}

This definition provides fairly easy discrimination
between transient and persistent sources. Usually the definition is
stricter (e.g., for LMXBs see Van Paradijs \& Verbunt 1984). In particular,
one often demands that the
difference between peak and quiescent flux is at least a factor 10$^3$.
However, this is difficult to test without observations with narrow-field
sensitive X-ray telescopes as is the case for many transients observed
in the 30-year history of space-borne X-ray astronomy.

The broad range of time scales allowed in the definition is chosen
to be able to include recent discoveries of bursting X-ray sources that
show no simultaneously detectable persistent emission.

\begin{figure*}
\centering
\includegraphics[width=1.0\linewidth]{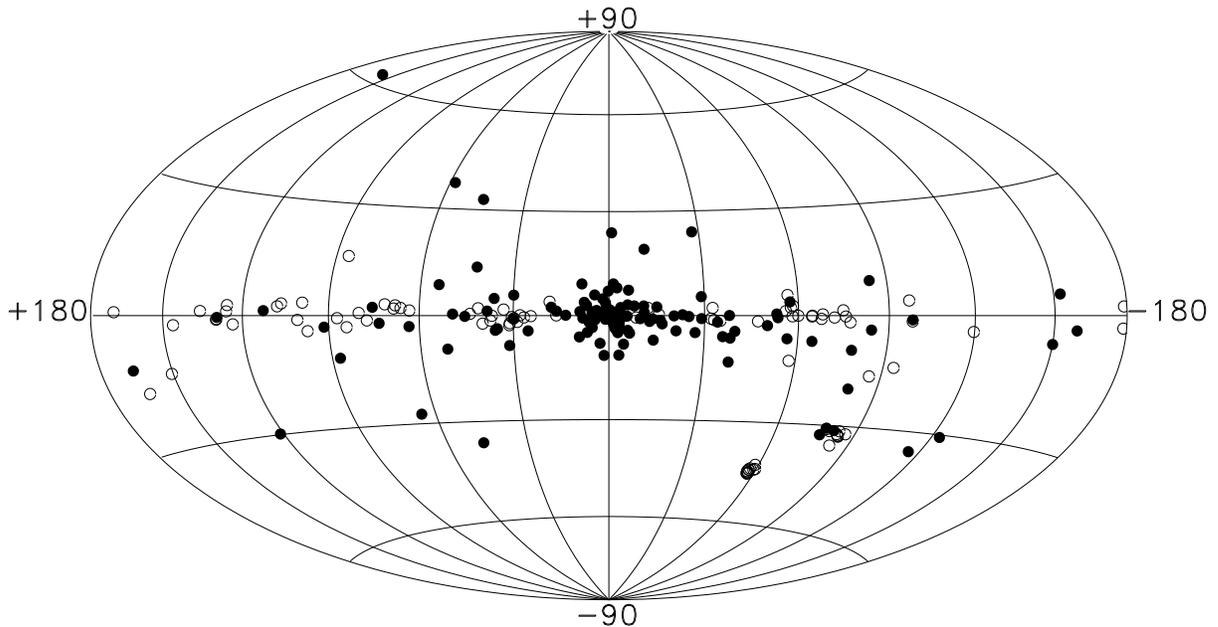}
\caption{Galactic distribution of 225 X-ray binaries (filled circles are LMXBs and
open circles HMXBs)\label{fig1}}
\end{figure*}

A variety of objects pertain to this definition. First of all,
the X-ray binaries in which a neutron star or black hole orbits a
non-degenerate star (for a review, see White et al. 1995). Furthermore,
it includes gamma-ray bursts,
nearby flare stars, pre-main sequence stars (with typical on-times
of a few hours), and RS CVn stars (up to a few days). All transients have
isotropic sky distributions except for the X-ray binaries (see
figure~\ref{fig1}). The high-mass X-ray
binaries, with companions to the compact stars of spectral types O and B,
consist of young objects that are concentrated along the galactic
plane. The low-mass X-ray binaries (LMXBs), with companions of type A and
later, are
older and are therefore typically found in globular clusters and 
in a concentration towards the galactic center. In fact, more than half
of all LMXBs are found within the central 20$^{\rm o}$
Monitoring programs of the galactic center region, therefore,
provide a unique opportunity to observe a large
fraction of the LMXB population in an unbiased manner.

In the remainder, we concentrate on LMXBs. 

\section{Past and current monitoring programs}

Six dedicated observation programs have been carried out to
monitor specifically the galactic center region for transient X-ray activity
above a few keV.
They are listed in table~\ref{tab1} together with the currently active RXTE
All-Sky Monitor (ASM) program as a reference.

The X-ray telescope (XRT) was operated during the 2nd Spacelab mission
in 1985 and was the first telescope to resolve an image of the galactic
center region above 2 keV (Skinner et al. 1987). The COded Mask Imaging
Spectrometer (COMIS) observed intermittently for 12 years
but most of the exposure was gathered before 1994 (In~'t~Zand 1992,
Emelyanov et al. 2000). ART-P on the Granat platform was operated for
a more limited time (Pavlinksy et al. 1994) than SIGMA on the same
platform. The SIGMA program distinguishes itself by the largest
exposure time of all programs and by being the first $\gamma$-ray imaging
device.

All instruments cover typical
field of views of 10$^{\rm o}$ combined with arcminute angular 
resolution and sensitivity in the 2-100 keV range (the same range
as for future INTEGRAL observations). 
All employed instruments except PCA/RXTE are coded aperture devices.

\noindent
\begin{center}
\begin{table*}
\caption[]{Overview of galactic center X-ray monitoring programs\label{tab1}.
The number of new sources is for within approximately 20$^{\rm o}$ from
the galactic center and includes confirmed sources weaker than 10~mCrab
(note that these are not included in table~\ref{tabtransients}.}
\begin{tabular}{lccclrccr}
\hline
Instrument & FOV & $\alpha^{\rm a}$  & S$^{\rm b}$ & yrs & Exposure & Number of & Number of & Refs.$^{\rm e}$\\
           & deg &            &  & & ksec & new sources & new bursters & \\
\hline
XRT/SL2                  &  6      &  3 & 0.5 & 85    & 25   &  6 &  1 & 1 \\
COMIS/Kvant/Mir$^{\rm c}$& 16      &  2 & 1   & 87-99 & 90   &  6 &  2 & 2,3 \\
ART-P/Granat             &  5      &  5 & 1   & 90-92 & 820  &  4 &  0 & 4 \\
SIGMA/Granat$^{\rm d}$   & 16      & 15 & 15  & 90-98 & 11000&  6 &  0 &5,6,7\\
WFC/BeppoSAX             & 40      &  5 & 10  & 96-   & 4000 & 11 & 17 & 8 \\
PCA/RXTE                 & 16      & 60 & 4   & 99-   & 8    &  8 &  1 & 9 \\
\hline
ASM/RXTE                 &         &  3 & 100 & 96-   & 990  &  2 &  0 & 10 \\
\hline
\end{tabular}

\noindent
\footnotesize
$^{\rm a}$Angular resolution in arcminutes (the source location accuracy
          is between 3 and 5 times better).\\
$^{\rm b}$Typical sensitivity in mCrab (2-10 keV).\\
$^{\rm c}$Also known as 'TTM'.\\
$^{\rm d}$SIGMA has no coverage below 35 keV, while the others do. The 
sensitivity is quoted for 35-150 keV\\
$^{\rm e}$References are:
1. Skinner et al. 1987;
2. In 't Zand 1992;
3. Emelyanov et al. 2000;
4. Pavlinsky et al 1994;
5. Vargas et al. 1996;
6. Vargas et al. 1997;
7. Kuznetsov et al. 1999;
8. This publication;
9. Markwardt et al. 1999 and priv. comm.;
10. Bradt et al. 1999.
\normalsize
\end{table*}
\end{center}

Currently, two programs are active: the BeppoSAX-WFC galactic bulge 
observations program and the RXTE-PCA bulge scan program. 

\subsection{BeppoSAX-WFC program}
\label{wfc}

\begin{figure}[h]
\centering
\includegraphics[width=1.0\linewidth]{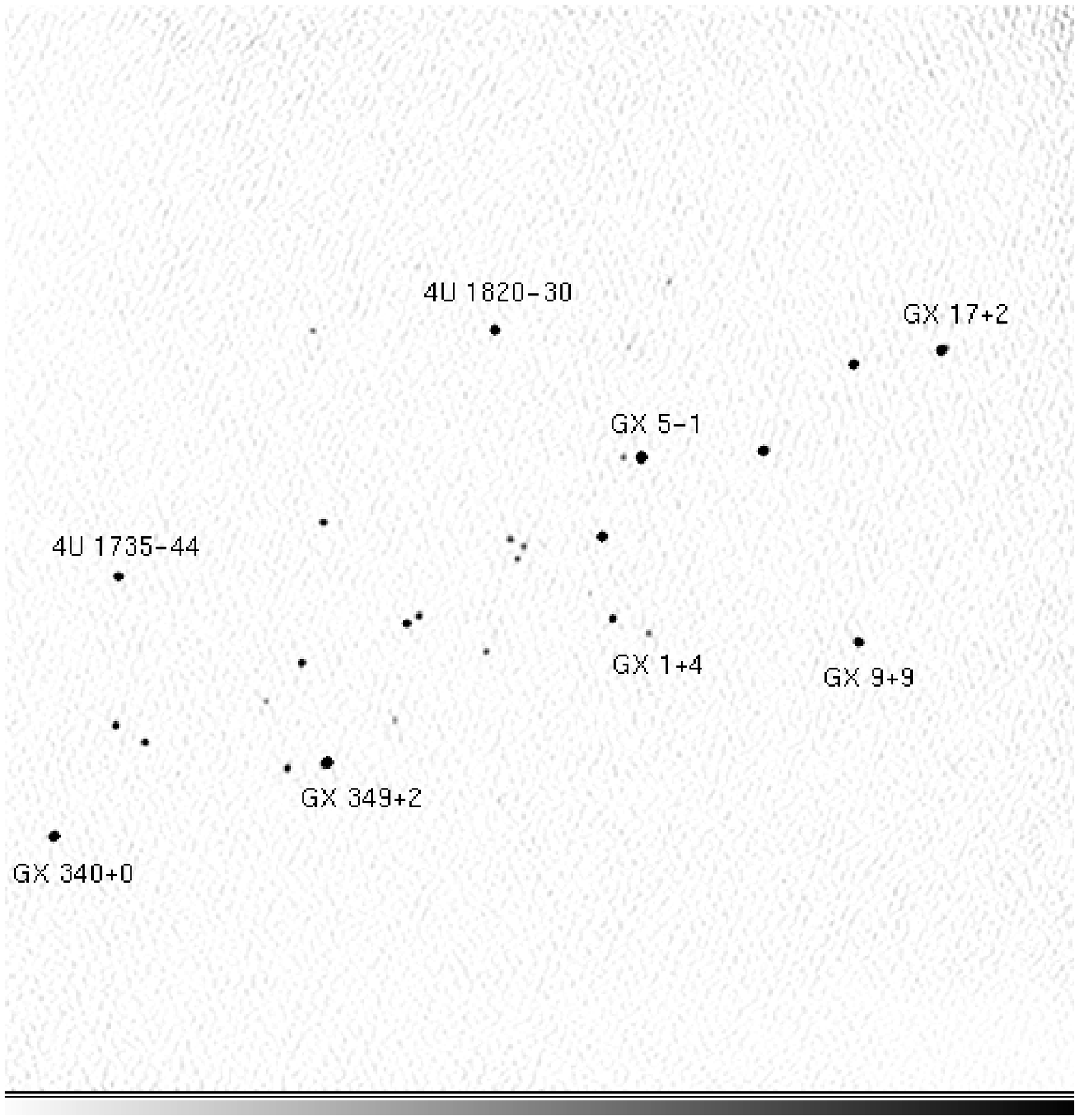}
\caption{Typical WFC image centered on the galactic center. Usually there
are about 30 active point sources above the detection limit\label{figwfc1}}
\end{figure}

The Wide Field Cameras (WFCs) on the BeppoSAX platform observe the
galactic bulge since Aug. 1996 weekly during each visibility
window allowed by the satellites operating constraints. These windows
are mid-February to mid-April and mid-August to mid-October of each
year. Up to the year 2000, nine campaigns were carried out amounting to
a total net exposure time of 4~Msec (or 10$^3$ hrs). This WFC program
takes 8\% of the complete BeppoSAX observation program and is thus
comparable, both in coverage and timing, to the INTEGRAL Galactic Center
Deep Survey core program. Following the census by
Van Paradijs (1995), the field of view of WFC encompasses more than
half of all known LMXBs. Thus, this program allows a unique monitoring
campaign of about 60 LMXBs with a large coverage and a
moderate sensitivity. This makes the program particularly suitable
to search for sub-hour flares or bursts.
Figure~\ref{figwfc1} shows a typical WFC image of the field.
The Principal Investigators are John Heise of SRON (Netherlands)
and Pietro Ubertini of IAS/CNR (Italy).

\begin{center}
\begin{table*}[t]
\caption[]{31 LMXB transients that were seen to be active in 1996-2000
within approximately 20 degrees from the galactic center. Many of the
parameter values for the bright transients were estimated from the
publicly available RXTE-ASM database (at URL {\tt http://xte.mit.edu}).
Other characteristics were obtained from the references listed, that are
sometimes arbitrary. Similar lists for transients before 1996 are given
in Chen et al. (1997).
\label{tabtransients}}
\begin{tabular}{lrrcll}
\hline
Name & Peak flux & Duration & Bursts? & Comment & References\\
     & mCrab     & days \\
\hline
GRO J1655-40     &  4500 & 200    & n & bhc      & Kuulkers et al. 2000 \\
X1658-298        &    30 & $>400$ & y &          & In 't Zand et al. 1999b \\
XTE J1709-267    &   200 & 100    & y &          & Cocchi et al. 1998 \\
XTE J1710-281    &    10 & $>600$ &   & eclipses & Markwardt et al. 1999 \\
2S 1711-339      &    50 & 150    & y &          & \\
SAX J1712.6-3739 &    50 & $>$100?& y &          & In 't Zand et al. 1999c\\
RX J1718-4029    & \multicolumn{2}{c}{\em burst only}&y&&Kaptein et al. 2000 \\
XTE J1723-376    &    80 & 70     & y &          & Marshall et al. 1999 \\
Rapid burster    &   300 & 70     & y &          & \\
GRS 1737-31      &    25 & 30     &   & bhc      & Cui et al. 1997 \\
GRS J1739-278    &   800 & 250    &   & jet, bhc & Vargas et al. 1997 \\
XTE J1739-285    &   200 & 40     &   &          & Markwardt et al. 2000a\\
KS 1741-293      &    30 & few    & y &          & In 't Zand et al. 1998a\\
GRS 1741.9-2853  &    70 & 10?    & y &          & Cocchi et al. 1999c\\
XTE J1743-363    &    15 & $>600$ &   &          & Markwardt et al. 1999\\
GRO J1744-28     &  2600 & 60     & y & type-II only? & \\
EXO 1745-248     &   600 & $>200$ & y & type-II? & Markwardt et al. 2000c\\
SAX J1747.0-2853 &   140 & 70     & y &          & Natalucci et al. 2000\\
GRS 1747-312     &    40 & 18     &   & eclipses & In 't Zand et al. 2000a\\
SAX J1748.9-2021 &    40 & 8      & y &          & In 't Zand et al. 1999a\\
XTE J1748-288    &   500 &15      &   & jet, bhc & Revnivtsev et al. 2000\\
SAX J1750.8-2900 &   100 & 6      & y &          & Natalucci et al. 1999a\\
SAX J1752.3-3138 & \multicolumn{2}{c}{\em burst only}&y&&Cocchi et al. 1999a\\
SAX J1753.5-2349 & \multicolumn{2}{c}{\em burst only}&y&&In 't Zand et al. 1998d\\
XTE J1755-324    &   150 & 40     &   & bhc      & Goldoni et al. 1999 \\
2S 1803-245      &   700 & 25     & y & jet      & Revnivtsev et al. 2000\\
SAX J1806.5-2215 & \multicolumn{2}{c}{\em burst only}&y&&In 't Zand et al. 1998d\\
SAX J1808.4-3658 &   100 & 18     & y & ms pulsar & In 't Zand et al., 1998c, 2000c\\
SAX J1810.8-2609 &    15 &  3     & y &          & Natalucci et al. 1999b\\
SAX J1818.6-1703 &   200 & 0.1    &   & uncertain LMXB         & In 't Zand et al. 1998a\\
SAX J1819.3-2525 & 12000 & 210    &   & jet, bhc & In 't Zand et al. 2000b \\
\hline
\end{tabular}
\end{table*}
\end{center}

\begin{figure}[h]
\centering
\rotatebox{90}{\includegraphics[width=1.05\linewidth,height=1.05\linewidth]{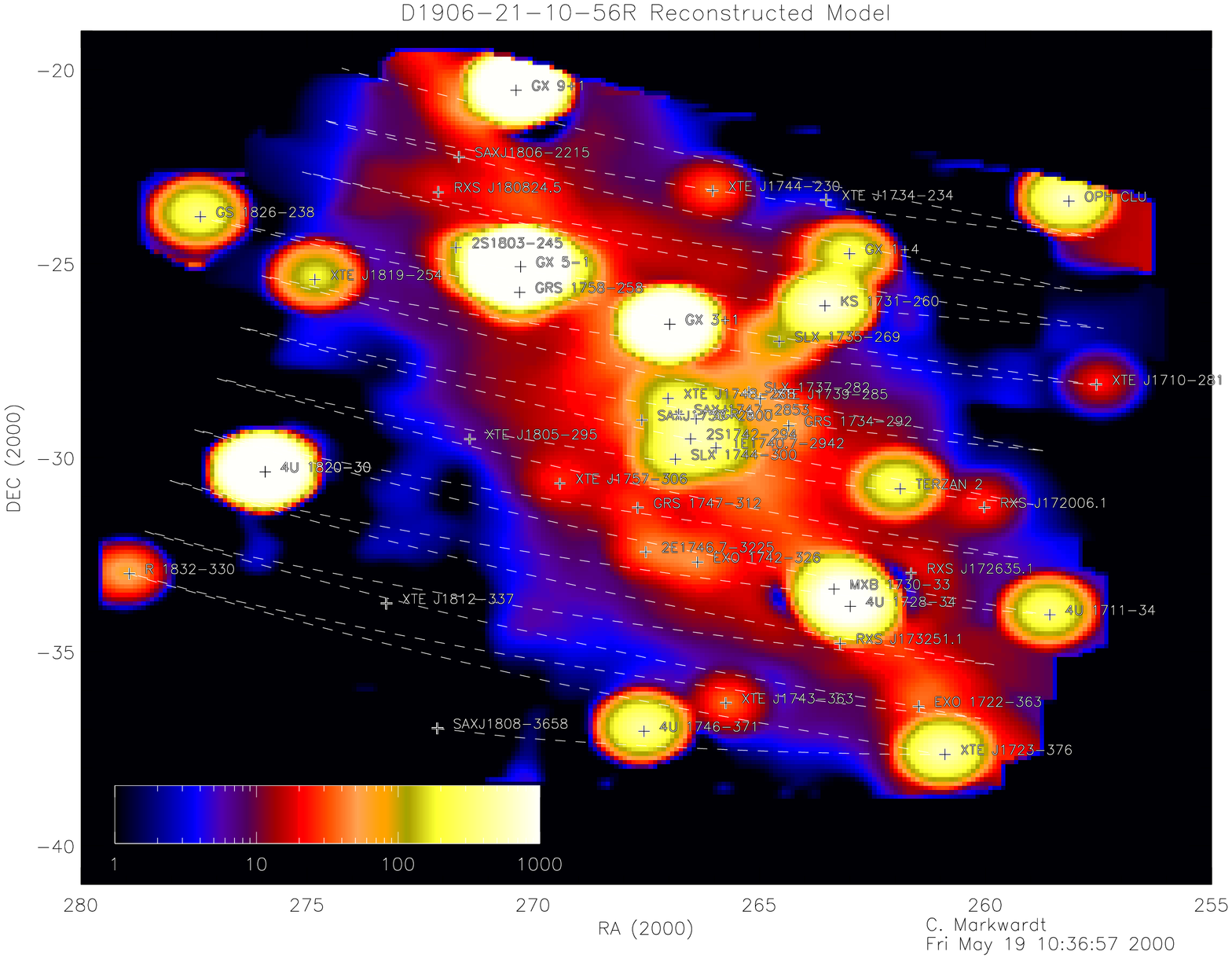}}
\caption{Reconstructed image of a typical PCA scan
observation made on Mar. 22, 1999.\label{figpca1} The dashed
curve shows the scan pattern for this particular observation }
\end{figure}

\subsection{RXTE-PCA program}

\begin{figure}[h]
\centering
\rotatebox{90}{\includegraphics[width=0.6\linewidth]{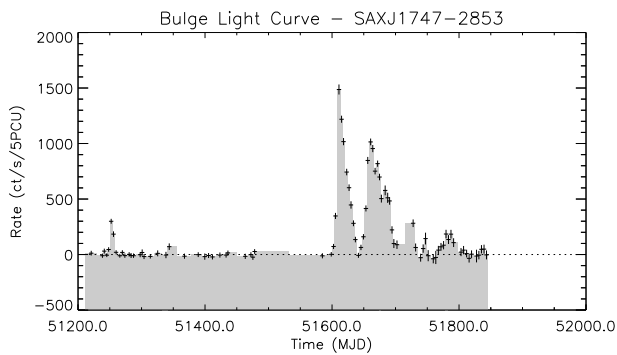}}
\caption{Light curve of SAX J1747.0-2853\label{fig1747} as obtained
with the RXTE-PCA bulge scan program since Feb. 1999 (Markwardt
et al. 2000b).}
\end{figure}

Since February 1999, scanning observations are carried
out with the Proportional Counter Array (PCA) on RXTE of a rectangular region
surrounding the galactic center, of approximately 16$^{\rm o}\times$18$^{\rm
o}$ and on a semi-weekly basis. The orientation of the field is such that
one diagonal is along the galactic plane. The scanning
pattern is zig-zag, alternately along each orientation. The slew
rate implies that each source in the field is in the field of view for
about 1 min, except for sources that are chosen to be at the end points of
the zig-zag pattern that have typical exposure times of 2 to 3 minutes.
Figure~\ref{figpca1} shows a typical reconstructed sky image of the field.
The sensitivity of these scan observations is about 1~mCrab over the whole
region. This is one to two orders of magnitude more sensitive than the 
observations with the All-Sky Monitor on 
RXTE. Thanks to the high frequency of the observations and
the high duty cycle throughout the year ($>80$\%), it is possible to uniformly 
sample outburst light curves of relatively faint transients. About 20
persistent sources are observed. Figure~\ref{fig1747} shows a typical
example of a lightcurve of a monitored source, namely the recurrent
transient SAX~J1747.0-2853. The PIs of this program are Craig
Markwardt and Jean Swank of NASA's Goddard Space Flight Center.

\section{Recent results}

If the X-ray binary catalog by Van Paradijs (1995,
last updated in 1992), is taken as a reference point, and when
the less clear identifications in that catalog are eliminated
(i.e., the 9 Einstein sources near the galactic
center, 3 unconfirmed weak KS sources, the anomalous X-ray pulsar 1E2259+587
and the white dwarf system CAL83), the total number of LMXBs known
is 110. The number
of new LMXB discoveries in the following years was 3 (1993), 3 (1994),
1 (1995), 4 (1996 and 1997), 6 (1998), 5 (1999) and 2 (2000, up to the
summer). Obviously, all of these new cases are transients. They 
expand the LMXB population by 25\%.

\begin{figure}[h]
\centering
\includegraphics[width=1.0\linewidth]{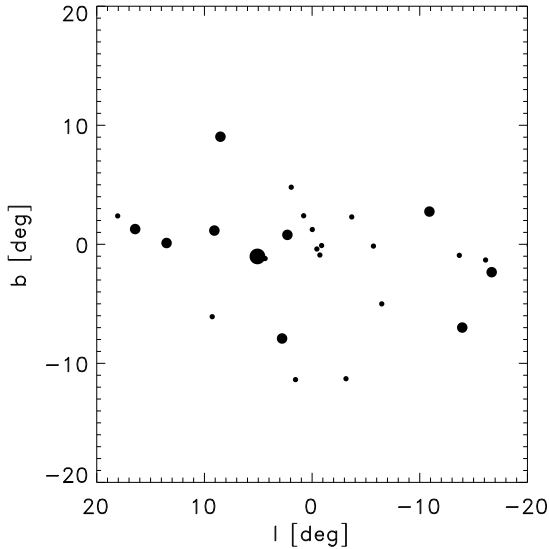}
\caption{Map of persistent LMXBs in galactic coordinates\label{figpers}.
Four symbol sizes code the average flux $f$: $f<0.1$ (smallest),
$0.1<f<1.0$, $1.0<f<10$, and $f>10$ (largest) Crab units in 2-10 keV.}
\end{figure}

\begin{figure}[h]
\centering
\includegraphics[width=1.0\linewidth]{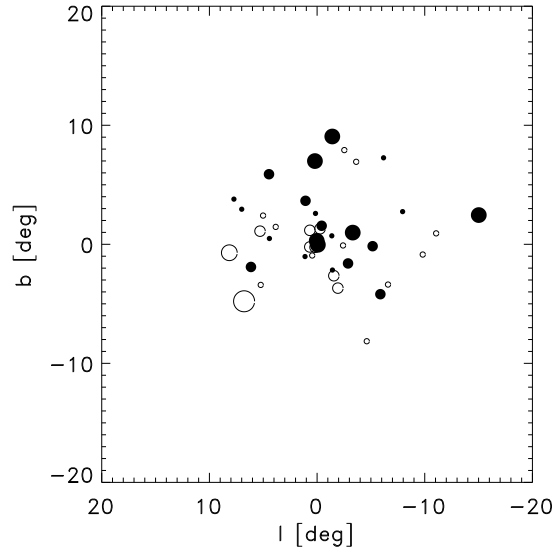}
\caption{Map of transient LMXBs in galactic coordinates\label{figtrans}.
The size coding is for peak flux and is the same as in figure~\ref{figpers}.
Open symbols denote discoveries after 1995.}
\end{figure}

Thirty-one LMXB transients were active during WFC and PCA observations in
1996-2000
(see table~\ref{tabtransients}). Interestingly, this is a large fraction of the
total number of 44 known transients in the field. Eight transients were
recurrent ones, with wait
times between outbursts measuring between 100 days (for the Rapid Burster) and
more than 20 years (e.g., MXB 1659-29). Eighteen of the 25 cases that are not
black hole candidates exhibited type-I X-ray bursts. Of the remaining 7 cases,
two are 'convincing' non-bursters because they are 'on' for long times
(XTE~J1710-281 and XTE~J1743-363).

\subsection{Galactic distribution}

In figures \ref{figpers} and \ref{figtrans} maps are
presented of all persistently bright and all transient
LMXBs, respectively, within 20$^{\rm o}$ from the galactic
center. There is a concentration towards the galactic center
that is stronger for the transients than 
for the persistently bright ones. This may partially be
due to selection effects. However, this selection effect does not
apply to WFC observations which were responsible for most new
discoveries. We derive that there is a strong suggestion for
a stronger concentration of transients.

\subsection{X-ray bursts}

\begin{table}[t]
\caption[]{List of the 35 bursters seen active with WFC in the field
(out of a total of 49 over the whole sky). The 17 sources
in italic are bursters first identified as such by the WFCs\label{tabbursters}.
XTE~J1723-376 was first recognized as a burster by RXTE.
This list was last updated in December 2000}

\begin{tabular}{ll}
\hline
Source           & References \\
\hline
%0512-401$^{\rm b}$              & \\
%0614+09                   & \\
%EXO 0748-676              & \\
%{\sl 2S 0918-549}         & \\
%{\sl 4U 1246-588}         & Piro et al. 1997 \\
%4U 1254-69                & \\
%{\sl SAX J1324.5-6313}    & in preparation \\
%4U 1608-522               & Smith et al. 1999 \\
%4U 1636-536               & \\
MXB 1658-298              & \\
4U 1702-429               & \\
4U 1705-440               & \\
{\sl XTE J1709-267}       & Cocchi et al. 1998 \\
{\sl 2S 1711-339}$^{\rm a}$     & Cornelisse, in preparation\\
{\sl SAX J1712.6-3739}    & Cocchi et al. 1999b \\
{\sl RX J1718.4-4029}     & Kaptein et al. 2000 \\
XTE J1723-376             & Marshall et al. 1999\\
1E 1724-3045              & Cocchi et al. 2000c\\
GX 354-0                  & \\
KS 1731-260               & Kuulkers, in preparation \\
Rapid burster$^{\rm b}$         & \\
{\sl SLX 1735-269}        & Bazzano et al. 1997 \\
4U 1735-44                & Cornelisse et al. 2000 \\
{\sl SLX 1737-282}        & in preparation \\
{\sl GRS 1741.9-2853}     & Cocchi et al. 1999c \\
KS 1741-293               & \\
A 1742-294                & \\
GRO J1744-28              & \\
SLX 1744-300              & \\
GX 3+1                    & in preparation \\
{\sl SAX J1747.0-2853}    & Natalucci et al. 2000 \\
{\sl SAX J1748.9-2021}$^{\rm b}$& In 't Zand et al. 1999a \\
{\sl SAX J1750.8-2900}    & Natalucci et al. 1999a \\
{\sl SAX J1752.4-3138}    & Cocchi et al. 1999a \\
{\sl SAX J1753.5-2349}$^{\rm a}$& In 't Zand et al. 1998d \\
{\sl 2S 1803-245}         & Muller et al. 1998 \\
{\sl SAX J1806.5-2215}    & In 't Zand et al. 1998d \\
{\sl SAX J1808.4-3658}    & In 't Zand et al. 1998b, 2000c \\
{\sl SAX J1810.6-2609}    & Natalucci et al. 1999b \\
4U 1812-12                & Cocchi et al. 2000a \\
GX 17+2                   & \\
%{\sl SAX J1818.7+1424}$^{\rm a}$& Cornelisse, in preparation \\
4U 1820-303$^{\rm b}$           & \\
{\sl GS 1826-24}          & Ubertini et al. 1999 \\
{\sl 1H 1832-33}$^{\rm b}$      & In 't Zand et al. 1998a \\
%Ser X-1                   & \\
%Aql X-1                   & \\
%4U 1915-05                & \\
%M15$^{\rm b}$                   & \\
\hline
\end{tabular}
\small
$^{\rm a}$uncertain type-I classification; $^{\rm b}$in globular cluster.
\normalsize
\end{table}

There are two types of X-ray bursts: type I and II. Both have
durations of order 1 minute and show fast-rise
exponential-decay time profiles. The difference is that type-I
bursts have few-keV black-body spectra with cooling in the tail
while type-II bursts do not cool and have spectra very
reminiscent to that of the emission outside bursts. Type-II
bursts are assumed to be due to instabilities in the accretion disk
(e.g., Lewin et al. 1993) while type-I bursts are recognized as thermonuclear 
flashes on surfaces of neutron stars. Currently, only two type-II
bursters are known (additionally, there is a tentative type-II
burster identification of the bright X-ray source in the globular cluster
Terzan 5; Markwardt et al. 2000c). Type-I bursts are powerful
diagnostics for the nature of the compact object: they eliminate
black holes or white dwarfs as candidates.

The WFCs have detected more than 2000
X-ray bursts in LMXBs, about 1500 of these are type-I bursts.
Table~\ref{tabbursters} presents a list of 35 bursters
that were seen active within 20$^{\rm o}$ from the galactic center.
This includes 17 new identifications of bursters. Bursts
were also detected from 14 sources outside this field, including 5 new ones.
Bursts were not detected by WFC from 16 non-listed and known bursters.
Therefore, WFC has seen 75\% of the burster population in an active state.

Particularly interesting bursting behavior was seen in GS~1826-24
where the bursts follow each other without exception during years
of observations quasi-periodically with a period of about 5.7~hr
(Ubertini et al. 1999).
This is in parellel to a very stable persistent flux which reflects a very
stable mass acretion rate by the neutron star. Changes on time scales
of years of the periodicity seem to go hand in hand with changes in
the persistent flux (Cocchi et al. 2000b). GS~1826-24 therefore provides
a unique testbed for theories on burst triggering mechanisms (e.g.,
Bildsten 2000).

Within the group of 49 persistent LMXBs, the fraction of bursters is
$0.6\pm0.2$ (note that this fraction can only grow). For transient LMXB, this
is $0.36\pm0.09$. The different fraction reflects the notion that the only
black
hole LMXBs found thus far are transient in nature. Within the GC field, the
brightest bursting LMXB transient is A1742-289. However, the identification of
A1742-289 with simultaneous bursting activity is not definite (Carpenter
1976 and Lewin et al. 1976, see also Maeda et al. 1996 and Kennea \& Skinner
1996). Apart from A1742-289, the brightest bursting LMXB transient is
2S~1803-245 (Muller et al., 1998 and Revnivtsev et al. 2000) with a peak
of 0.7 Crab units. All others have peak fluxes at least a factor of 2 less.
The brightest bursting persistent LMXB in the field is GX~17+2 with an
average flux of 0.6 Crab units.

\subsection{SAX J1808.4-3658}

Arguably, SAX~J1808.4-3658 is the most intriguing X-ray transient
found since the microquasars GRS~1915+105 and GRO~J1655-40.
Discovered in 1996 by the WFCs as a 0.1~Crab X-ray transient
(In 't Zand et al. 1998c), it was first measured with the more
sensitive PCA in
a second outburst in early 1998 when a pulsar signal was found with
a period of 2.5~ms (Wijnands \& Van der Klis 1998). This
represents the first and sofar only discovery of an accreting
millisecond pulsar. Such
systems were long sought after because LMXBs are thought to
be the precursor of millisecond radio pulsars (e.g., Bhattacharya \&
Van den Heuvel 1991).

\begin{figure}[t]
\centering
\includegraphics[width=1.0\linewidth]{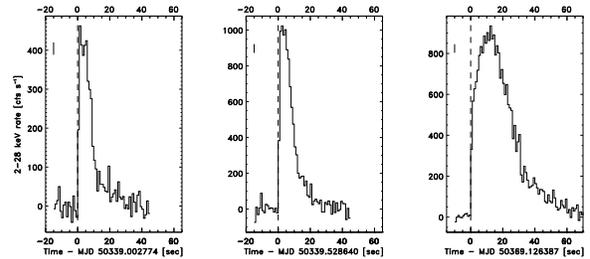}
\caption{Time profiles of 3 bursts from SAX~J1808.4-3658\label{fig1808A}}
\end{figure}

There is indirect evidence for millisecond neutron star rotation
periods in other LMXBs. Transient millisecond oscillations have
been found in RXTE-PCA data during (parts of) X-ray bursts in 10 LMXBs
(4U~1728-34, 4U~1636-53, 4U~1702-429, KS~1731-260, Aql~X-1,
X1658-298, 4U~1608-52, 4U~1916-05, the Rapid Burster and a source close
to galactic center). Since the
asymptotic frequencies of most of these oscillations are reproducable
over years, the suggestion is strong
to interpret these as due to the neutron star rotation
(Strohmayer 1999). However, this is not a settled issue.
SAX~J1808.4-3658 could help resolve this issue because this
is the only burster
for which a rotation period has been independently determined.
So far, bursts were only detected with the WFCs.

The WFC data from the first detected outburst from SAX~J1808.4-3658 
has recently been revisited, employing higher data coverage and
a more mature calibration (In~'t~Zand et al. 2000c). This resulted
in the detection of
a third burst which was 50\% brighter than the other 2 bursts
already published by In~'t~Zand et al. (1998c), though with a similar
bolometric peak flux, see figure~\ref{fig1808A}.
This is the brightest type-I burst detected by WFC among about 1500 cases,
and is perhaps the only case for which meaningful searches for burst
oscillations
in the moderately sensitive WFC data are possible (note that the photon
collecting area for WFC is about 40 times smaller than for
RXTE's prime instrument).

\begin{figure}[h]
\centering
\includegraphics[width=1.0\linewidth]{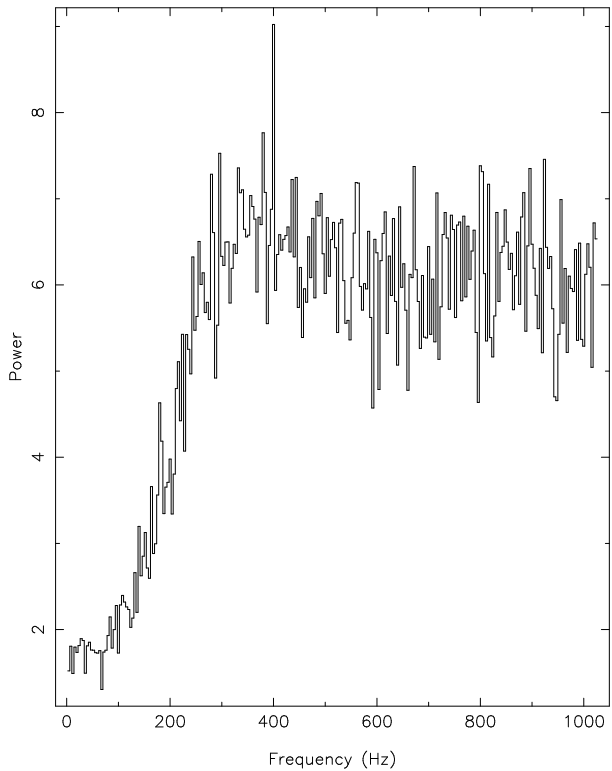}
\caption{Power density spectrum of first 20 sec of 3rd burst from
SAX~J1808.4-3658\label{fig1808B}}
\end{figure}

A number of attempts to find a signal at 400 Hz resulted in the intriguing
power density spectrum presented in figure~\ref{fig1808B}. This
is the average spectrum over 80 intervals of 0.25~sec starting
at the onset of the burst. Therefore, the frequency resolution
is rather broad at 4~Hz. At the time of this workshop, the significance
of this signal is being assessed and the indications are that it is
on the edge of being significant.

\subsection{Is there a separate class of faint transients?}

Heise et al. (1999) reports that seven LMXB transients
discovered with the WFCs in this field
were faint and that a large subset (six) consists of X-ray bursters.
They propose that these transients make up an as yet
unrecognized
subclass of LMXBs. We here note eight more transients in the field:

\begin{itemize}
\item SAX J1712.6-3739 is a transient with a measured peak flux
of 32 mCrab (In 't Zand et al. 1999c). It exhibited a burst (Cocchi et al.
1999b);
\item SAX J1752.3-3138 was seen only during a burst (Cocchi et al. 1999a).
The persistent emission was below 6 mCrab at the time of the burst;
\item SAX J1818.6-1703 was a brief transient that was active for
a few hours at a peak flux of about 0.2 Crab (In 't Zand et al. 1998a).
It did not exhibit bursts. Technically speaking
this is not necessarily a LMXB but we include this source because it
is in this field;
\item SAX J1819.3-2525 (In 't Zand et al. 2000b) was mostly faint
except for a few hours when it was seen to be very bright 
(12 Crab units; Smith et al. 1999). It is now proven to be
a black hole system at a distance between 7.4 and 12.3 kpc (Orosz et al.
2001). Despite a companion of early spectral type, this transient
is more characteristic for a LMXB transient than for a HMXB transient
(presumably because this is a semi-detached binary without a 
supergiant companion);
\item XTE J1710-281 is a faint transient which is active since at least 1998
and showing a lot of variability below 10 mCrab, including eclipses (Markwardt
et al. 1999);
\item XTE J1723-376 was a moderately bright transient in early 1999
 when it exhibited type-I bursts (Marshall et al. 1999);
\item XTE J1739-285 is a fairly bright transient that stayed on for at least
 a few weeks (Markwardt et al. 2000a);
\item XTE J1743-363 is a faint transient showing variable activity below
about 15 mCrab since 1999 (Markwardt et al. 1999).
\end{itemize}
In total, fifteen faint LMXB transients have been detected, with 9 or 10 of
them bursting. These are not necessarily
short transients, as observations have shown of for example SAX~J1747.0-2853
(figure~\ref{fig1747}) and SAX~J1712.6-3739 (Cocchi et al., in preparation)
which all are active for at least 200 days.
This broadens the characteristics of the proposed subclass of faint LMXB transients
considerably, and we conclude that such a subclass is perhaps not as
strictly defined as initially thought. Nevertheless, the BeppoSAX and RXTE
observations do show that
there is a large fraction of likely LMXB transients that do not reach high peak
luminosities. Perhaps this is related to a special nature of the
companion star. For instance, the companion star
in SAX~J1808.4-3658 appears to
be of very low mass ($\leq0.1$~M$_\odot$, Chakrabarty \& Morgan 1998). 
Unfortunately, identifications of optical counterparts of these transients
are scarce and difficult in this field, and measurements of orbits through
pulsar timing have so far only been possible in SAX~J1808.4-3658.

\section{Conclusions}

In the last 4 years, monitoring programs by BeppoSAX-WFC and RXTE-PCA on the
Galactic center field have helped extend the LMXB population by tens of percents.
Most of the new transients are fainter than the classical X-ray novae, and are type-I
X-ray bursters. The frequency of LMXB transients brighter than approximately 10 mCrab in
our galaxy is $12\pm4$ per year. The percentage of bursters, and therefore
confirmed neutron star systems, is about 60\% versus about 40\% for all LMXB transients.
The higher burster percentage among fainter transients may be
a selection effect because low accretion rates are more favorable toward triggering
thermonuclear flashes on neutron star surfaces, if present. Therefore, this
measurement does not necessarily indicate a higher neutron star to black
hole ratio among fainter transients.

The newly found LMXB transients comply fairly well with results from an
archival study of 66 outbursts from 24 X-ray novae by Chen et al. (1997).
The peak luminosity distribution for X-ray novae was shown to
range from 0.003 to 5 times the Eddington luminosity, peaking at 0.2
$L_{\rm edd}$ which is equivalent to roughly a few tenths of a Crab unit if at a
distance equal to that of the galactic center.

\section*{Acknowledgments}

The author thanks the organization of this workshop for the
invitation to give this presentation, John Heise and Pietro
Ubertini for allowing a discussion of the WFC program, Craig
Markwardt and Jean Swank for allowing a discussion of the PCA
bulge program and providing figures \ref{figpca1} and
\ref{fig1747}, and Angela Bazzano and John Heise for careful
reading of the manuscript.

% The following bibliography was produced with
%   \bibliographystyle{aa}
%   \bibliography{esapub}
% The results are inserted directly here to simplify
% the demonstration.

\end{document}